\documentclass{article}

\usepackage{arxiv}

\usepackage[utf8]{inputenc} % allow utf-8 input
\usepackage[T1]{fontenc}    % use 8-bit T1 fonts
\usepackage{hyperref}       % hyperlinks
\usepackage{url}            % simple URL typesetting
\usepackage{booktabs}       % professional-quality tables
\usepackage{amsfonts}       % blackboard math symbols
\usepackage{nicefrac}       % compact symbols for 1/2, etc.
\usepackage{microtype}      % microtypography
\usepackage{lipsum}		% Can be removed after putting your text content
\usepackage{graphicx}
\usepackage{natbib}
\usepackage{doi}
\usepackage{newtxtext}
\usepackage{newtxmath}
\usepackage{authblk}

\title{Optimal waveform for fast synchronization of airfoil wakes}

\author[ ]{Vedasri Godavarthi$^1$\thanks{Corresponding author: vedasrig@g.ucla.edu}}
\author[2]{Yoji Kawamura}
\author[1]{Kunihiko Taira}
\affil[1]{Department of Mechanical and Aerospace Engineering, University of California, Los Angeles, CA 90095, USA}
\affil[2]{Center for Mathematical Science and Advanced Technology, Japan Agency for Marine-Earth Science and Technology, Yokohama 236-0001, Japan}

\hypersetup{
pdftitle={A template for the arxiv style},
pdfsubject={q-bio.NC, q-bio.QM},
pdfauthor={David S.~Hippocampus, Elias D.~Striatum},
pdfkeywords={First keyword, Second keyword, More},
}

\begin{document}
\maketitle

\begin{abstract}
	We obtain an optimal actuation waveform for fast synchronization of periodic airfoil wakes through the phase reduction approach. Using the phase reduction approach for periodic wake flows, the spatial sensitivity fields with respect to the phase of the vortex shedding are obtained. The phase sensitivity fields can uncover the synchronization properties in the presence of periodic actuation. This study seeks a periodic actuation waveform using phase-based analysis to minimize the time for synchronization to modify the wake-shedding frequency of NACA0012 airfoil wakes. This fast synchronization waveform is obtained theoretically from the phase sensitivity function by casting an optimization problem. The obtained optimal actuation waveform becomes increasingly non-sinusoidal for higher angles of attack. Actuation based on the obtained waveform achieves rapid synchronization within as low as two vortex shedding cycles irrespective of the forcing frequency whereas traditional sinusoidal actuation requires $\mathcal{O}(10)$ shedding cycles. Further, we analyze the influence of actuation frequency on the vortex shedding and the aerodynamic coefficients using force-element analysis. The present analysis provides an efficient way to modify the vortex lock-on properties in a transient manner with applications to fluid-structure interactions and unsteady flow control.
\end{abstract}

% keywords can be removed
%\keywords{First keyword \and Second keyword \and More}

	\section{Introduction}
	\label{sec:intro}
	
	Unsteady periodic fluid flows are common in nature and engineering settings including vortex shedding over flapping wings, bluff bodies and airfoils. Modifying the vortex shedding behavior of wake flows is of high relevance to developing efficient engineering systems. However, controlling such flows is challenging owing to their periodically varying base states \citep{colonius2011control}. For the time-periodic base state, the timing of actuation becomes important. For this purpose, it is necessary to characterize the perturbation dynamics with respect to the time-periodic base state, which can be achieved using a phase reduction technique \citep{winfree1967biological,kuramoto1984chemical}. The phase reduction approach expresses the perturbation dynamics using a single scalar phase variable. Recently, it has been used for studying periodic flows to reveal the phase sensitivity fields \citep{kawamura2013collective,kawamura2015phase,khodkar2020phase,kawamura2022adjoint,loe2021phase,iima2021phase}, synchronization characteristics to external forcing \citep{taira2018phase,khodkar2020phase,khodkar2021phase,skene2022phase} and flow control \citep{nair2021phase, loe2023controlling} in a computationally inexpensive way. 

 Examining synchronization properties for periodic wakes can offer insights to modify the vortex shedding behavior and has several applications in unsteady flow control and fluid-structure interactions. Control of vortex shedding of wake flows has direct implications towards modifying the aerodynamic characteristics, reduction of structural vibration and noise emissions. Synchronization control has been studied in the context of vortex-induced vibrations for bluff-body wakes \citep{feng2010circular, konstantinidis2016vortex}. In addition, actuating a flow by taking advantage of synchronization can be efficient in enhancing the aerodynamic performance \citep{pastoor2008feedback,joe2011feedback,wang2018enhancement,asztalos2021modeling}. Further, hydrodynamic synchronization is shown to result in efficient swimming in microscale swimmers at lower Reynolds number \citep{golestanian2011hydrodynamic, kawamura2018phase}. Hence, it is beneficial to analyse the parameters that result in optimal synchronization in fluid flows. While most synchronization studies for fluid flows have characterized this asymptotic synchronization process to external sinusoidal actuation \citep{taira2018phase,khodkar2020phase,herrmann2020modeling,giannenas2022harmonic}, it is often desirable to modify the vortex shedding as quickly as possible for flow control to take effect. This study considers the fast synchronization of wakes to an external forcing signal for wake flows. 

The concept of fast synchronization has been studied in biology to promote the rapid adjustment of the biological clock to jet lag and facilitate treatments for cardiac arrhythmias \citep{guevara1982phase,granada2009achieve}. In the context of biological and simpler oscillatory systems, \cite{zlotnik2013optimal} and \cite{takata2021fast} applied the phase reduction approach to analytically obtain the optimal waveform for fast synchronization, maximizing the synchronization speed to external periodic forcing. In dynamical systems, synchronization is also referred to as entrainment \citep{strogatz1994chaos}.

\begin{figure}

   \includegraphics[width=1.1\textwidth]{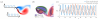}
	\caption{ Fast synchronization analysis of flow over NACA0012 airfoil at $\alpha=55^\circ$ and $\text{Re}=100$. $(a)$ Periodic actuation using the optimal waveform. $(b)$ Comparison of synchronized flowfield for a forcing frequency $\Omega_f = 0.9\Omega_n$ with the baseline vorticity field $\boldsymbol{\omega}$. $(c)$ Lift coefficient $C_L(t)$ when actuated with the fast synchronization and sinusoidal waveforms.} 
\label{fig:fig01_overview}
\end{figure}

In this study, we apply such a phase reduction approach to perform the phase-based fast synchronization for NACA0012 airfoil wakes at post-stall angles of attack with leading- and trailing-edge actuation. The overview of the fast synchronization analysis is shown in figure~\ref{fig:fig01_overview}. We analytically find the optimal actuation waveforms, and the airfoil wake flows are actuated numerically using the optimal and sinusoidal waveforms at various forcing frequencies. The respective synchronization speeds are compared to validate the theoretical results. Further, we investigate the influence of actuation on the flow fields and the lift coefficients. The paper is organized as follows. The phase-based description and the framework to obtain the waveform for fast synchronization are presented in Section \ref{sec:methodology}. The current approach is demonstrated with an example of NACA0012 airfoil wakes in Section \ref{sec:results}. Conclusions are offered in Section \ref{sec:conclusions}.
 %Further, results are validated numerically by actuation of the airfoil with theoretically identified optimal waveform at different frequencies. We analyze the effect of actuation on the flowfields and the aerodynamic coefficients using the lift force element theory. The overview of the present work is shown in Fig. \ref{fig:fig01_overview} with NACA0012 and $\alpha=55^\circ$ as an example. First, the phase description is made through the lift coefficient plane $C_L-\dot{C}_L$ and the phase sensitivity fields with respect to velocity perturbations are obtained through the adjoint-based phase reduction approach. The waveform for fast entrainment is obtained theoretically based on these phase sensitivity fields. These waveforms are compared with the sinusoidal waveforms corresponding to the same forcing direction. To validate these results numerically, we actuate the NACA0012 airfoil at $\alpha=55^\circ$ with the optimal waveform at frequencies different than natural frequency. The speed of entrainment is then compared with actuating the airfoil using a sinusoidal waveform. 

% 	\begin{figure}
%	\centering
%    \includegraphics[width=\textwidth]{figs/fig01_overview.pdf}
%	\caption{Overview of fast entrainment analysis. (a) Phase description of airfoil wakes using $C_L-\dot{C}_L$ plane. (b) Spatial phase sensitivity fields with respect to velocity perturbations obtained used an adjoint-based phase reduction approach. (c) Optimal waveform for fast entrainment obtained from phase sensitivity fields theoretically and comparison with a sinusoidal waveform. (d) Implementation of pointwise actuation numerically at optimal locations near leading and trailing edges. (e) Comparison of entrained flowfield at a forcing frequency $\Omega_f = 0.9\Omega_n$ with baseline vorticity. (f) Comparison of $C_L(t)$ when actuated with fast entrainment and sinusoidal waveform.} 
%	\label{fig:fig01_overview}
%\end{figure}
 
	\section{Fast synchronization analysis through the phase-reduction approach}
	\label{sec:methodology}
To obtain the optimal actuation waveform for fast synchronization for periodic fluid flows, we use phase reduction analysis \citep{taira2018phase, kawamura2022adjoint}. We identify the phase sensitivity fields that encode the effect of timing of actuation and then analytically solve an optimization problem to obtain the synchronization waveform in terms of the phase sensitivity function.
 \subsection{Phase reduction approach}
We consider incompressible time-periodic fluid flows governed by the Navier-Stokes equations $ \dot{\boldsymbol{q}}=\mathcal{N}(\boldsymbol{q}(\boldsymbol{x},t))$, where $\boldsymbol{q}$ is the flow state. These equations are given by
 \begin{align}
    \frac{\partial \boldsymbol{q}}{\partial t} &= -\boldsymbol{q}\cdot \nabla \boldsymbol{q}-\nabla p +\frac{1}{\text{Re}} \nabla^2 \boldsymbol{q}, \nonumber \\
    \nabla \cdot \boldsymbol{q} &=0,
    \label{eq:N-S}
 \end{align}
 \noindent where $\text{Re}$ is the Reynolds number and $p$ is the pressure. For a time-periodic flow $\boldsymbol{q}_0(\boldsymbol{x},t)$, it satisfies $\boldsymbol{q}_0(\boldsymbol{x},t+T)=\boldsymbol{q}_0(\boldsymbol{x},t)$, where $T$ is the time period of the limit cycle and $\Omega_n = 2\pi/T $ is the natural frequency of the system. Here, we define a phase $\theta$ such that 
 \begin{equation}
 \dot{\theta}=\Omega_n,\: \theta \in\, [0,2\pi).
 \end{equation}

 \noindent With the definition of $\theta$, we can identify the full state vector of the limit cycle solution $\boldsymbol{q}_0(\boldsymbol{x},\theta)$ at every $\theta$. Given a stable limit cycle solution, with the frequency of the limit cycle being $\Omega_n$, the phase in the vicinity of the limit cycle can be described using the generalized phase variable $\Theta(\boldsymbol{q}(\boldsymbol{x},t))$. Thus, the generalized phase dynamics is described as 
 \begin{equation}
    \dot{\Theta}(\boldsymbol{q}) = \int_{\mathcal{D}}\nabla_{\boldsymbol{q}} \Theta(\boldsymbol{q}) \cdot \dot{\boldsymbol{q}} d \boldsymbol{x} = \int_{\mathcal{D}}\nabla_{\boldsymbol{q}}\Theta(\boldsymbol{q}) \cdot \mathcal{N}(\boldsymbol{q}) d \boldsymbol{x} = \Omega_n.
 \end{equation}
\noindent Leveraging the phase dynamics, we can derive the phase response to sufficiently small perturbations,
 \begin{equation}
 \dot{\boldsymbol{q}}=\mathcal{N}(\boldsymbol{q})+\epsilon \boldsymbol{F}(\boldsymbol{x},t),
 \label{eq:NS_force}
 \end{equation}
which provides the corresponding change to phase dynamics as
\begin{align}
\dot{\theta}(t)&=\dot{\Theta}(\boldsymbol{q}) = \int_{\mathcal{D}} \nabla_{\boldsymbol{q}}\Theta(\boldsymbol{q}) \cdot \dot{\boldsymbol{q}} d \boldsymbol{x} = \int_{\mathcal{D}} \nabla_{\boldsymbol{q}} \Theta(\boldsymbol{q}) \cdot \left[\mathcal{N}(\boldsymbol{q}(\boldsymbol{x},t))+\epsilon \boldsymbol{F}(\boldsymbol{x},t)\right] d \boldsymbol{x} \nonumber\\
&\approx  \Omega_n + \epsilon\int_{\mathcal{D}}\boldsymbol{Z}(\boldsymbol{x},\theta)\cdot \boldsymbol{F}(\boldsymbol{x},t) d \boldsymbol{x}.
\label{eq:phase_dyn}
 \end{align}

\noindent Here, $\boldsymbol{Z}(\boldsymbol{x},\theta)=\nabla_{\boldsymbol{q}} \Theta(\boldsymbol{q})|_{\boldsymbol{q}=\boldsymbol{q}_0(\boldsymbol{x},\theta)}$ is the spatial phase sensitivity field as it quantifies the phase response of the system to any given small perturbation and $\mathcal{D}$ is the considered spatial domain. This spatial phase sensitivity field can be obtained using either a direct impulse-based method \citep{taira2018phase,khodkar2020phase,nair2021phase,loe2021phase}, or an adjoint-based approach \citep{kawamura2013collective,kawamura2015phase,kawamura2022adjoint}, or a Jacobian-free approach \citep{iima2021phase}. We use the adjoint-based phase reduction framework to obtain the phase sensitivity fields in the present study as we can obtain the high-fidelity $\boldsymbol{Z}(\boldsymbol{x},\theta)$ in all the flow variables by solving a single pair of forward and adjoint simulations.
%The direct impulse-based method quantifies the phase response to an impulse-based perturbation at all the phases and grid points. 

\subsection{Adjoint-based approach for phase sensitivity fields}
We utilize the adjoint-based formulation to find the phase sensitivity fields $\boldsymbol{Z}$ for time-periodic wakes. Let us consider the dynamics of a small perturbation $\boldsymbol{q}^\prime(\boldsymbol{x},\theta,t)$ by linearizing the Navier-Stokes equations about the periodic base state $\boldsymbol{q}_0(\boldsymbol{x},\theta)$. The perturbation dynamics is given by $\dot{\boldsymbol{q}}^\prime = \mathcal{L}(\boldsymbol{x},\theta)\boldsymbol{q}^\prime$, where $\mathcal{L}(\boldsymbol{x},\theta)$ is the linearized Navier--Stokes operator.

To obtain the phase dynamics of the dominant limit cycle oscillation, we consider Floquet eigenfunction $\boldsymbol{Q}$ and adjoint eigenfunction $\boldsymbol{Q}^*$ corresponding to the zero eigenvalue. This Floquet-zero eigenvalue corresponds to the phase degree of freedom for the stable limit cycle dynamics. Thus phase dynamics is obtained by projecting the perturbed dynamics in equation~\ref{eq:NS_force} on the adjoint eigenfunction as
\begin{align}
    \dot{\theta}(t) &= \int_{\mathcal{D}} \left[ \boldsymbol{Q}^*(\boldsymbol{x},\theta) \cdot \mathcal{N}(\boldsymbol{q})+\epsilon  \boldsymbol{Q}^*(\boldsymbol{x},\theta) \cdot \boldsymbol{F}(\boldsymbol{x},t) \right] d\boldsymbol{x} \nonumber \\
    &\approx \Omega_n + \epsilon \int_{\mathcal{D}} \boldsymbol{Q}^*(\boldsymbol{x},\theta) \cdot \boldsymbol{F}(\boldsymbol{x},t)  d\boldsymbol{x},
    \label{eq:adjoint_phase}
\end{align}

\begin{figure}
\centering
   \includegraphics[width=1.05\textwidth]{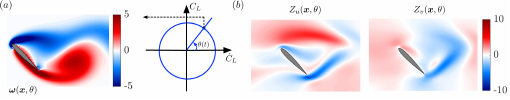}
	\caption{ An overview of the phase reduction approach for flow over a NACA0012 airfoil at $\alpha=45^\circ$ and $\textrm{Re}=100$. $(a)$ Definition of phase based on the lift coefficient, $C_L-\dot{C}_L$ plane. $(b)$ Instantaneous spatial phase sensitivity fields $Z_u$ and $Z_v$ with respect to the perturbations in velocity fields.} 
\label{fig:fig02_phase}
\end{figure}

\noindent We note that the norm of $\boldsymbol{Q^*}$ is arbitrary and the normalization condition of $\boldsymbol{Q^*}$ is appropriately chosen to satisfy $\int_{\mathcal{D}} \left[ \boldsymbol{Q}^*(\boldsymbol{x},\theta) \cdot \mathcal{N}(\boldsymbol{q}) \right] d\boldsymbol{x}=\Omega_n$ for all $\theta$. More details on the properties of $\boldsymbol{Q}^*$ are given in our earlier work \citep{kawamura2022adjoint}. Here, by comparing equations~\ref{eq:adjoint_phase} $\&$ \ref{eq:phase_dyn}, for perturbations in the form of velocity, we obtain $\boldsymbol{Z}(\boldsymbol{x},\theta) = \boldsymbol{Q}^*(\boldsymbol{x},\theta)$. Hence, phase sensitivity is the adjoint-zero eigenfunction of the linearized Navier-Stokes operator. The spatial phase sensitivity fields can be obtained by solving the dynamics, which in two dimensions is governed by the linearized adjoint equations of
\begin{align}
    \frac{\partial }{\partial t}\boldsymbol{Q}^*(\boldsymbol{x},-\Omega_n t) &= -U^* \nabla u -V^* \nabla v + \boldsymbol{q} \cdot \nabla \boldsymbol{Q}^*-\nabla P^* +\frac{1}{\text{Re}}\nabla ^2 \boldsymbol{Q}^*, \nonumber \\
    \nabla \cdot \boldsymbol{Q}^* &=0,
    \label{eq:adjoint_NS}
\end{align}
 where $\boldsymbol{Q}^*=\left(U^*,\, V^*\right)$ and $\boldsymbol{q}=\left(u,\,v\right)$. Thus, the phase sensitivity fields with respect to perturbations in the velocity field are obtained by seeking a periodic solution for equation~\ref{eq:N-S} and solving the system of adjoint equations (equation~\ref{eq:adjoint_NS}). Since, the adjoint equations are analogous to the Navier--Stokes equations, the same numerical scheme can be used to solve them. An overview of the phase description for airfoil wakes is shown in figure~\ref{fig:fig02_phase}. The phase is defined based on the lift coefficient $C_L-\dot{C}_L$ plane, where $\theta=0,\pi$ correspond to $\textrm{mean }C_L$, $\theta = \pi/2$ corresponds to maximum $C_L$ and $\theta=3\pi/2$ corresponds to minimum $C_L$ \citep{taira2018phase}.

\subsection{Synchronization analysis to external periodic forcing}

 Establishing this oscillator dynamics enables us to study the synchronization characteristics of the system to an external periodic forcing signal with a frequency $\Omega_f$ different from the wake shedding frequency $\Omega_n$. We introduce a localized periodic forcing at location $\boldsymbol{x}_0$ at a forcing frequency $\Omega_f$. The spatial profile of forcing is given by a Dirac delta function $h(\boldsymbol{x})=\delta(\boldsymbol{x}-\boldsymbol{x}_0)$. Using equation~\ref{eq:phase_dyn}, the governing phase dynamics becomes
 \begin{equation}
\dot{\theta}(t) = \Omega_n + \epsilon\boldsymbol{\zeta}(\theta)\cdot \boldsymbol{f}(\Omega_f t),
 \end{equation}
 where $\boldsymbol{F}(\boldsymbol{x},t)=\boldsymbol{f}(\Omega_f t)h(\boldsymbol{x})$ and the local phase sensitivity function is given by $\boldsymbol{\zeta}(\theta) = \int_{\mathcal{D}} \boldsymbol{Z}(\boldsymbol{x},\theta) h(\boldsymbol{x}) d\boldsymbol{x}=\boldsymbol{Z}(\boldsymbol{x}_0,\theta)$. To characterize the synchronization of the system to external forcing, we consider the relative phase $\phi(t)$ between the phase of the system $\theta(t)$ and that of the forcing signal $\Omega_f t$ as $ \phi(t) = \theta(t) - \Omega_f t$. The dynamics of the relative phase is provided as
 \begin{equation}
     \dot{\phi}(t) = \Omega_n - \Omega_f + \epsilon\boldsymbol{\zeta}(\phi(t)+\Omega_f t)\cdot \boldsymbol{f}(\Omega_f t) = \Delta\Omega +\epsilon\boldsymbol{\zeta}(\phi(t)+\Omega_f t)\cdot \boldsymbol{f}(\Omega_f t),
 \end{equation}
 \noindent where $\Delta \Omega = \Omega_n-\Omega_f$. The asymptotic behaviour of relative phase dynamics can be obtained by averaging over a period of forcing \citep{kuramoto1984chemical, ermentrout1991multiple},
 \begin{equation}
      \dot{\phi}(t) = \Delta \Omega +\epsilon\Gamma(\phi),
      \label{eq:relative_phase}
 \end{equation}
 where 
 \begin{equation}
    \Gamma(\phi)=\frac{1}{2\pi}\int\limits_{0}^{2\pi}\boldsymbol{\zeta}(\phi+\psi)\cdot \boldsymbol{f}(\psi) d\psi
 \end{equation}
 is the phase coupling function and $\Delta \Omega = \Omega_n - \Omega_f$.
Synchronization occurs if the relative phase becomes a constant, i.e., $\dot{\phi}\rightarrow 0$. Hence, the synchronization condition is given as
\begin{equation}
    \epsilon \min\limits_{\phi}\Gamma(\phi) \leq -\Delta \Omega \leq  \epsilon  \max\limits_{\phi}\Gamma(\phi).
\end{equation}
The synchronization condition determines the forcing frequency required to synchronize the dynamics to the external actuation based on the phase coupling function. 

We aim to identify the optimal periodic actuation to synchronize the system to a forcing frequency as quickly as possible. Hence, the rate of convergence of $\phi$ to a fixed point $\phi_*$ should be maximized to satisfy
\begin{equation}
    \dot{\phi}_*=\Delta \Omega + \epsilon \Gamma(\phi_*)=0.
    \label{eq:entrainment}
\end{equation}
Therefore, we can formulate an optimization problem to maximize $|\dot{\phi}|$, which occurs when $-\epsilon \Gamma^\prime(\phi_*)$ is large. Here $-\Gamma^\prime(\phi_*)$ is the synchronization speed $S$. The cost function $\mathcal{J}$ is therefore formulated as
\begin{equation}
    \mathcal{J}(\boldsymbol{f}) = -\Gamma^\prime(\phi_*) - \lambda\left(\langle  \boldsymbol{f}\cdot \boldsymbol{f}\rangle-1\right) - \mu\left(\Delta \Omega+\epsilon\Gamma(\phi_*)\right),
    \label{eq:optimization}
\end{equation}
where $\lambda$ and $\mu$ are Lagrangian multipliers and $\langle \cdot \rangle=\frac{1}{2\pi}\int\limits_0^{2\pi}\left(\cdot\right) d\theta$. The first term corresponds to maximizing the synchronization speed, the second term constrains the energy of actuation, and the third term directly follows from equation~\ref{eq:entrainment}. 

Since the synchronization is independent of the initial phase, without loss of generality, we consider the fixed point, $\phi_*=0$. This optimization can be solved analytically using the calculus of variations \citep{zlotnik2013optimal}.
The optimal waveform for fast synchronization can then be derived as
\begin{equation}
     \boldsymbol{f}(\theta;\Delta \Omega/\epsilon) = -\frac{\boldsymbol{\zeta}^\prime(\theta)}{2\lambda} - \frac{(\Delta \Omega /\epsilon) \boldsymbol{\zeta}(\theta)}{\langle\boldsymbol{\zeta}\cdot \boldsymbol{\zeta}\rangle},\:\: \lambda = \frac{1}{2}\sqrt{\frac{\langle \boldsymbol{\zeta}^\prime \cdot \boldsymbol{\zeta}^\prime\rangle }{1-\frac{(\Delta \Omega/\epsilon)^2}{ \langle \boldsymbol{\zeta} \cdot \boldsymbol{\zeta} \rangle}}}.
    \label{eq:optimal_waveform}
\end{equation}
Hence, once we compute the local phase sensitivity function $\boldsymbol{\zeta}(\theta)$, the optimal waveform for fast synchronization can be analytically found using equation~\ref{eq:optimal_waveform} for various $\Omega_f$ and $\epsilon$. The optimal speed of synchronization is characterized by then computing $-\Gamma^\prime(0)$ using the optimal waveform given by equation~\ref{eq:optimal_waveform}. Even though the current formulation to obtain the optimal waveform is defined for a single pointwise actuation, this directly extends to the case with multiple pointwise actuators or spatially distributed actuators following from equation \ref{eq:phase_dyn}. This is reflected as a change in the inner product $\langle \cdot \rangle$ in equations \ref{eq:optimization} and \ref{eq:optimal_waveform}, where the inner product would be computed as an integration for the multiple local phase sensitivities and actuation waveforms. Next, we uncover these optimal waveforms for the airfoil wakes using the local phase sensitivity functions and assess their performance for fast synchronization.

	\section{Phase synchronization analysis of airfoil wakes} 	\label{sec:results}
 \subsection{Computational set-up}
This study considers the two-dimensional incompressible laminar flow over NACA0012 airfoils at angles of attack, $\alpha=35^\circ,\,45^\circ$ and $55^\circ$ and chord-based Reynolds number of $\text{Re}=U_\infty c/\nu=100$, where $U_\infty,\,c$ and $\nu$ are the free-stream velocity, airfoil chord length and kinematic viscosity, respectively. The flow dynamics is governed by incompressible Navier-Stokes equations \ref{eq:N-S} and the obtained flow fields present with periodic vortex shedding \citep{kawamura2022adjoint}. The actuation in equation \ref{eq:NS_force} is introduced as a localized force with the form $ \boldsymbol{F}(\boldsymbol{x},\Omega_f t) = \boldsymbol{f}(\Omega_f t)\delta(\boldsymbol{x}-\boldsymbol{x}_0)$, where $\boldsymbol{x}_0$ is a forcing location. The Dirac delta function is approximated with a three-cell discrete delta function \citep{roma1999adaptive}. 

The periodic flows over the airfoil are computed numerically through the immersed boundary projection method \citep{taira2007immersed, kajishima2016computational}. For the numerical simulation, we consider a computational domain $\mathcal{D}=(x/c,y/c)\in [-16,16]\times[-30,30]$. The quarter-chord of the airfoil is placed at the origin. 
The smallest grid size is set to $\Delta x_{\min}/c = 0.02$, and the time step is chosen to be $\Delta t = 0.005$. The present computational setup has been validated and is the same as that used in \cite{kawamura2022adjoint}. The same computational setup is used for adjoint simulations of the phase sensitivity fields.

\begin{figure}
\centering
	\includegraphics[width=0.8\textwidth]{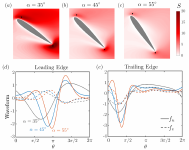}
	\caption{$(a)$-$(c)$ Synchronization speed $S$ around a NACA0012 airfoil at $\alpha=35^\circ,45^\circ$ and $55^\circ$. The black dots indicate local maxima. $(d)$-$(e)$ Theoretical optimal waveforms for fast synchronization with pointwise forcing at the leading and trailing edges for $\alpha=35^\circ,45^\circ$ and $55^\circ$. } 
	\label{fig:fig02_effect_aoa}
\end{figure}
\subsection{Synchronization analysis for airfoil wakes}

 The spatial phase sensitivity fields with respect to the streamwise and transverse velocity components $Z_u$ and $Z_v$ for NACA0012 airfoils at $\alpha = 35^\circ,\,45^\circ$ and $55^\circ$ are obtained through the adjoint-based approach described in Section 2.2. Using the obtained spatial phase sensitivity fields, we can compute the optimal waveform for fast synchronization at each grid point as per equation~\ref{eq:optimal_waveform}, which is then used to obtain the optimal synchronization speed at each grid point. We investigate the effect of the angle of attack on the synchronization speed and waveforms of NACA0012 airfoil wakes, as shown in figure \ref{fig:fig02_effect_aoa}. We consider the case when the forcing frequency $\Omega_f =\Omega_n$ and $\Delta \Omega =0$. It follows from equation~\ref{eq:optimal_waveform} that the optimal actuation waveform at each point is proportional to the corresponding derivative of the local phase sensitivity function $\boldsymbol{\zeta}^\prime(\theta)$. The spatial distributions of synchronization speed $S$ around the airfoil found using the optimal waveform for $\alpha = 35^\circ, 45^\circ$ and $55^\circ$ are depicted in figures~\ref{fig:fig02_effect_aoa}$(a)$-$(c)$. As the angle of attack increases, the overall magnitude of the synchronization speed decreases, indicating an increased difficulty in synchronization for higher post-stall angles of attack. With an increase in $\alpha$, we observe stronger and larger leading- and trailing-edge vortex structures. To achieve synchronization with external forcing, the vortex formation time and the length scale have to be modified. This therefore becomes challenging with higher $\alpha$, which is reflected in reduced synchronization speed. Further, we also note that the white region around the airfoil corresponds to a small optimal synchronization speed, indicating that, irrespective of the actuation energy, these spatial locations are not conducive for flow modification. This means that the actuation effort must penetrate the outside of the boundary layer. 
 
 We also observed that, for all $\alpha$, the local maxima in the synchronization speed are attained near the leading and trailing edges, suggesting them as optimal actuation locations for synchronization (indicated as black dots). The leading and trailing edges are the most sensitive regions since they are specific regions in the flow field with high curvature. Further, the regions with high synchronization speed become more compact and are concentrated at the leading and trailing edges with an increase in $\alpha$ due to the earlier flow separation and the concentration of gradients at the leading and trailing edges at higher $\alpha$. Even though we considered three angles of attack, we expect this trend in synchronization speed and optimal waveform to hold true for much higher angles of attack. 
 
 For the NACA0012 airfoil at $Re=100$, we observe a steady wake until $\alpha \approx 20^\circ$. This results in a constant lift coefficient and, hence, an ill-defined phase. Further, as $\alpha\rightarrow 90^\circ$, we approach a zero mean lift coefficient. However, for $30^\circ \leq \alpha \leq 90^\circ$, we observe periodic vortex shedding and we can leverage the optimal waveform analysis. We expect a similar trend in the optimal waveform and in the synchronization speed with an increase in the angle of attack. An increase in the angle of attack results in the formation of stronger leading- and trailing-edge vortices, thereby increasing the difficulty in synchronization and the reduction in synchronization speed. The asymmetry in the vortex formation and roll-up between the leading- and trailing-edge vortices also increases with most $\alpha$, resulting in a non-sinusoidal optimal waveform. However, for $\alpha \rightarrow 90^\circ$, we approach a symmetric bluff-body vortex shedding. Hence, overall, the optimal waveform outperforms the synchronization speed of a sinusoidal waveform at most higher angles of attack. It is noteworthy that, due to the difficulty in synchronization at higher angles of attack, we will require a larger actuation effort to synchronize the wake to a different frequency.
 
 The optimal actuation waveforms in the $x$ and $y$ velocity directions, at the leading and trailing edges for various $\alpha$, are shown in figures~\ref{fig:fig02_effect_aoa}$(d)$-$(e)$. As $\alpha$ increases, the optimal waveform becomes increasingly non-sinusoidal, due to the asymmetry in the vortex formation and shedding process near the leading and trailing edges at higher angles of attack. Further, the optimal waveform at the trailing edge at higher angles of attack suggests a smaller time duration where actuation is significant (for $0<\theta<3\pi/4$ in figure~\ref{fig:fig02_effect_aoa}(e)), in comparison with the leading-edge optimal waveform. This is in line with flow physics, as we observe a more compact and stronger vortex roll-up at the trailing edge when compared to the vortex formation at the leading edge. We would like to point out that these optimal waveforms are obtained by independently maximizing the synchronization speed using the respective local phase sensitivity functions. We can also obtain the optimal waveforms at the leading and trailing edges by optimizing the synchronization speed using the local phase sensitivity functions simultaneously. For this present study, both these cases lead to similar results with minimal modification, where simultaneous optimization results in the same waveforms but with more actuation energy at the trailing edge than at the leading edge. This difference should be carefully considered for more complex flow fields when using multiple actuation locations.
 %We compare this optimal waveform with a sinusoidal waveform of the same average actuation direction as shown in Fig.\ref{fig:fig01_overview}(c). The theoretical entrainment speed computed using the optimal waveform increases by 2.5 times in comparison with sinusoidal forcing for $\alpha=55^\circ$.
\begin{figure}
\centering
	\includegraphics[width=0.75\textwidth]{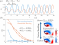}
	\caption{Numerical results for synchronization analysis of the NACA0012 airfoil at $\alpha=45^\circ$.~$(a)$ Changes in $C_L(t)$ at a forcing frequency $\Omega_f=1.05\Omega_n$. $(b)$ Synchronization time using sinusoidal and optimal waveforms at different forcing frequencies. $(c)$ Comparison of instantaneous vorticity fields for forcing frequencies, $\Omega_f=0.95\Omega_n,\Omega_f=1.05\Omega_n$ with the unperturbed vorticity field.} 
	\label{fig:fig03_convergence}
\end{figure}

Next, we numerically validate the synchronization analysis by introducing actuation at the optimal actuation locations near the leading and trailing edges (as shown in figure~\ref{fig:fig02_effect_aoa})(d)-(e)). Here, we consider the optimal waveform and a sinusoidal waveform with the same averaged actuation direction at different forcing frequencies.  We present the numerical results at $\alpha=45^\circ$ as a representative case. The numerical results of synchronization for a forcing frequency within $5\%$ of the natural frequency are shown in figure~\ref{fig:fig03_convergence}. Here, we choose an actuation amplitude of $\epsilon=0.1$ to achieve synchronization for laminar flows at higher angles of attack.

To assess the synchronization speed, we consider cycle-to-cycle variations of the $C_L$ coefficient and measure the inter-peak phase difference $\Delta \theta_k$ for each cycle as shown in figure~\ref{fig:fig03_convergence}$(a)$. The optimal waveform actuation achieves synchronization in two shedding cycles, in comparison to $\mathcal{O}(10)$ shedding cycles for the sinusoidal waveform (see figure~\ref{fig:fig03_convergence}$(b)$) for different forcing frequencies. Since the optimal waveform is based on the phase sensitivity function, it can efficiently identify the ``when'' and ``how'' to efficiently synchronize the system to an external forcing signal, thus achieving fast synchronization. The effect of actuation frequency on flow physics is examined using the instantaneous vorticity fields of synchronized and unperturbed in figure~\ref{fig:fig03_convergence}$(c)$. We observe streamwise elongation of the leading and trailing edge vortices for lower frequency actuation, $\Omega_f = 0.95\Omega_n$ when compared with the unperturbed case. On the other hand, we observe more compact leading and trailing edge vortices for higher actuation frequencies, $\Omega_f = 1.05\Omega_n$. Hence, the modification of vortex shedding frequency through optimal waveform actuation is achieved by modifying the vortex formation length scale near the leading and trailing edges. Thus, the phase-sensitivity-based optimal waveform deviates from the sinusoidal waveform to target more actuation energy at the right time to achieve rapid flow modification.

To further examine the effect of the present actuation over the lift coefficients, let us monitor the force elements \citep{chang1992potential}. Force element theory enables us to identify the flow structures responsible for lift generation. We compute an auxiliary potential function $\phi_L$ that satisfies the Laplace equation $\nabla^2 \phi_L=0$, with the boundary condition $-\boldsymbol{n}\cdot \nabla\phi_L=\boldsymbol{n}\cdot \boldsymbol{e}_y$ on the airfoil surface, where $\boldsymbol{e}_y$ is the unit vector in the lift direction. The lift force is obtained by taking the inner product of $\nabla\phi_L$ with the momentum equation and integrating with $\mathcal{D}$ in two dimensions as
\begin{equation}
    F_L = \int\limits_{\mathcal{D}} \boldsymbol{\omega} \times \boldsymbol{u} \cdot \nabla \phi_L dD + \frac{1}{\text{Re}}\int\limits_{\partial \mathcal{D}} \boldsymbol{\omega} \times \boldsymbol{n} \cdot (\nabla \phi_L + \boldsymbol{e}_y) dl,
\end{equation} where the first term denotes the surface integral and the second term denotes the line integral on the airfoil surface. The first integrand herein referred to the lift element $L_E$, and is used to monitor the effect of vortical structures on the lift force.

\begin{figure}
\centering
	\includegraphics[width=1.05\textwidth]{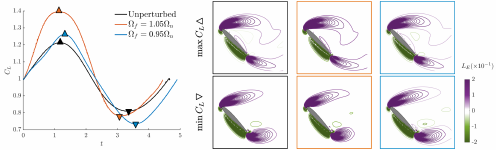}
	\caption{Characterization of $C_L$ for the NACA0012 airfoil at $\alpha = 45^\circ$ and $\text{Re}=100$. Variation of $C_L$ when actuated with forcing frequencies $\Omega_f=1.05\Omega_n$ and $0.95\Omega_n$. Instantaneous lift force elements $L_E$ are shown for the actuated and unperturbed cases at $\max C_L (\Delta)$ and $\min C_L (\nabla)$.} 
	\label{fig:fig04_liftelement}
\end{figure}

The lift coefficient $C_L$ for a vortex shedding period for the unperturbed and the actuation frequencies $\Omega_f=0.95\Omega_n$ and $1.05 \Omega_n$ are shown in figure~\ref{fig:fig04_liftelement}. The snapshots are shown corresponding to the unperturbed flow fields (black box), synchronized flow fields at $\Omega_f=0.95\Omega_n$ (blue box) and $\Omega_f =1.05\Omega_n$ (red box) at $\max C_L (\Delta)$ and $\min C_L (\nabla)$. Owing to the actuation, we notice a significant change in $C_L$ compared to the unperturbed case for both frequencies, especially for $\Omega_f=1.05\Omega_n$. For a $5\%$ increase in frequency ($\Omega_f=1.05\Omega_n$), we observe a $17\%$ increase in $\max C_L$ and a $8\%$ increase in mean $C_L$ compared with the unperturbed case. However, we do note that a similar amount of actuation is introduced to the flowfield. It is noteworthy that the swift modification of the shedding timing is achieved by the lift increases for high-frequency actuation. We further analyse the wake with the lift elements $L_E$ $(\Delta)$ for unperturbed (black box) and high-frequency actuation (red box) as shown in figure~\ref{fig:fig04_liftelement}. We observe a strong compact positive $L_E$ near the leading and trailing edges. This suggests that the increased strength and compactness of the vortex increases the local circulation, and thereby the lift force \citep{eldredge2019leading}.

We now consider the low-frequency actuation ($\Omega_f=0.95\Omega_n$), where we do not observe a significant change in mean $C_L$ in comparison with the unperturbed case. In contrast to the high-frequency actuation, the optimal waveform actuation achieves a reduction in the wake shedding frequency through a reduction in $\min C_L$. The lift force elements $L_E$ corresponding to this case (blue box,$\nabla$) show a streamwise-elongated positive force element effectively pushing away the shear layer from the airfoil surface, thereby reducing the overall lift force. Overall, through the lift element theory, we identified that high-frequency actuation using the optimal waveform results in compact vortices at the leading and trailing edges, and the lower wake shedding frequency is achieved by streamwise elongation of the vortices at the leading and trailing edges. Through a high-frequency actuation using the optimal waveform, a transient increase in lift is observed, albeit with a considerable actuation effort. By demonstrating the effectiveness of optimal waveform analysis for $\alpha=45^\circ$, we show the potential of this method for the analysis of a wide range of periodic fluid flows and their control in a transient manner.

\section{Conclusions}
\label{sec:conclusions}

We presented a theoretical framework to find an optimal actuation waveform for maximizing the synchronization speed for periodic fluid flows. This was demonstrated for periodic post-stall airfoil wakes using localized forcing.  We leveraged the phase reduction approach to identify the sensitivity with respect to the vortex shedding phases, thereby identifying the right time and direction of actuation for efficient synchronization. The optimal actuation waveform for fast synchronization departs from a sinusoidal waveform for higher angles of attack. We showed that the optimal waveform significantly outperforms the sinusoidal waveform in terms of synchronization speed. We further identified that the modification of wake shedding frequency is achieved by the elongation of vortical structures, whereas synchronization to a higher frequency is achieved by compacting vortical structures near the leading and trailing edges. The present study based on phase reduction with an optimal waveform approach holds potential to develop transient flow control strategies that produce a quick response.
	
\section*{Acknowledgments}
	\label{sec:acknowledgments}
V.G and K.T acknowledge the support from the US Air Force Office of Scientific Research (Grants: FA9550-21-1-0178, FA9550-22-1-0013) and the US National Science Foundation (Grant: 2129639). Y.K. acknowledges financial support from JSPS (Japan) KAKENHI Grant Numbers JP20K03797, JP18H03205 and JP17H03279. Y.K. also acknowledges support from Earth Simulator JAMSTEC Proposed Project.
	
	\section*{Declaration of interest}
	\label{sec:doi}
	The authors report no conflict of interest.

	%\bibliography{taira_refs}
 \bibliographystyle{unsrtnat}
	\bibliography{refs}

\begin{thebibliography}{35}
\providecommand{\natexlab}[1]{#1}
\providecommand{\url}[1]{\texttt{#1}}
\expandafter\ifx\csname urlstyle\endcsname\relax
  \providecommand{\doi}[1]{doi: #1}\else
  \providecommand{\doi}{doi: \begingroup \urlstyle{rm}\Url}\fi

\bibitem[Colonius and Williams(2011)]{colonius2011control}
T.~Colonius and D.~R. Williams.
\newblock Control of vortex shedding on two-and three-dimensional aerofoils.
\newblock \emph{Philos. Trans. Royal Soc.}, 369\penalty0 (1940):\penalty0 1525,
  2011.

\bibitem[Winfree(1967)]{winfree1967biological}
A.~T. Winfree.
\newblock Biological rhythms and the behavior of populations of coupled
  oscillators.
\newblock \emph{J. Theo. Bio.}, 16\penalty0 (1):\penalty0 15--42, 1967.

\bibitem[Kuramoto(1984)]{kuramoto1984chemical}
Y.~Kuramoto.
\newblock \emph{Chemical Oscillations, Waves, and Turbulence}.
\newblock Springer, 1984.

\bibitem[Kawamura and Nakao(2013)]{kawamura2013collective}
Y.~Kawamura and H.~Nakao.
\newblock Collective phase description of oscillatory convection.
\newblock \emph{Chaos: An Inter. J. Nonlin. Sci.}, 23\penalty0 (4):\penalty0
  043129, 2013.

\bibitem[Kawamura and Nakao(2015)]{kawamura2015phase}
Y.~Kawamura and H.~Nakao.
\newblock Phase description of oscillatory convection with a spatially
  translational mode.
\newblock \emph{Phys. D: Nonlin. Phen.}, 295:\penalty0 11--29, 2015.

\bibitem[Khodkar and Taira(2020)]{khodkar2020phase}
M.~A. Khodkar and K.~Taira.
\newblock Phase-synchronization properties of laminar cylinder wake for
  periodic external forcings.
\newblock \emph{J. Fluid Mech.}, 904:\penalty0 R1, 2020.

\bibitem[Kawamura et~al.(2022)Kawamura, Godavarthi, and
  Taira]{kawamura2022adjoint}
Y.~Kawamura, V.~Godavarthi, and K.~Taira.
\newblock Adjoint-based phase reduction analysis of incompressible periodic
  flows.
\newblock \emph{Phys. Rev. Fluids}, 7\penalty0 (10):\penalty0 104401, 2022.

\bibitem[Loe et~al.(2021)Loe, Nakao, Jimbo, and Kotani]{loe2021phase}
I.~A. Loe, H.~Nakao, Y.~Jimbo, and K.~Kotani.
\newblock Phase-reduction for synchronization of oscillating flow by
  perturbation on surrounding structure.
\newblock \emph{J. Fluid Mech.}, 911:\penalty0 R2, 2021.

\bibitem[Iima(2021)]{iima2021phase}
M.~Iima.
\newblock Phase reduction technique on a target region.
\newblock \emph{Phys. Rev. E}, 103\penalty0 (5):\penalty0 053303, 2021.

\bibitem[Taira and Nakao(2018)]{taira2018phase}
K.~Taira and H.~Nakao.
\newblock Phase-response analysis of synchronization for periodic flows.
\newblock \emph{J. Fluid Mech.}, 846:\penalty0 R2, 2018.

\bibitem[Khodkar et~al.(2021)Khodkar, Klamo, and Taira]{khodkar2021phase}
M.~A. Khodkar, J.~T. Klamo, and K.~Taira.
\newblock Phase-locking of laminar wake to periodic vibrations of a circular
  cylinder.
\newblock \emph{Phys. Rev. Fluids}, 6\penalty0 (3):\penalty0 034401, 2021.

\bibitem[Skene and Taira(2022)]{skene2022phase}
C.~S. Skene and K.~Taira.
\newblock Phase-reduction analysis of periodic thermoacoustic oscillations in a
  {Rijke} tube.
\newblock \emph{J. Fluid Mech.}, 933:\penalty0 A35, 2022.

\bibitem[Nair et~al.(2021)Nair, Taira, Brunton, and Brunton]{nair2021phase}
A.~G. Nair, K.~Taira, B.~W. Brunton, and S.~L. Brunton.
\newblock Phase-based control of periodic flows.
\newblock \emph{J. Fluid Mech.}, 927:\penalty0 A30, 2021.

\bibitem[Loe et~al.(2023)Loe, Zheng, Kotani, and Jimbo]{loe2023controlling}
I.~A. Loe, T.~Zheng, K.~Kotani, and Y.~Jimbo.
\newblock Controlling fluidic oscillator flow dynamics by elastic structure
  vibration.
\newblock \emph{Sci. Rep.}, 13\penalty0 (1):\penalty0 8852, 2023.

\bibitem[Feng and Wang(2010)]{feng2010circular}
L.~H. Feng and J.~J. Wang.
\newblock Circular cylinder vortex-synchronization control with a synthetic jet
  positioned at the rear stagnation point.
\newblock \emph{J. Fluid Mech.}, 662:\penalty0 232--259, 2010.

\bibitem[Konstantinidis and Bouris(2016)]{konstantinidis2016vortex}
E.~Konstantinidis and D.~Bouris.
\newblock Vortex synchronization in the cylinder wake due to harmonic and
  non-harmonic perturbations.
\newblock \emph{J. Fluid Mech.}, 804:\penalty0 248--277, 2016.

\bibitem[Pastoor et~al.(2008)Pastoor, Henning, Noack, King, and
  Tadmor]{pastoor2008feedback}
M.~Pastoor, L.~Henning, B.~R. Noack, R.~King, and G.~Tadmor.
\newblock Feedback shear layer control for bluff body drag reduction.
\newblock \emph{J. Fluid Mech.}, 608:\penalty0 161--196, 2008.

\bibitem[Joe et~al.(2011)Joe, Colonius, and MacMynowski]{joe2011feedback}
W.~T. Joe, T.~Colonius, and D.~G. MacMynowski.
\newblock Feedback control of vortex shedding from an inclined flat plate.
\newblock \emph{Theo. Comp. Fluid Dyn.}, 25:\penalty0 221--232, 2011.

\bibitem[Wang and Tang(2018)]{wang2018enhancement}
C.~Wang and H.~Tang.
\newblock Enhancement of aerodynamic performance of a heaving airfoil using
  synthetic-jet based active flow control.
\newblock \emph{Bioinsp. Biomim.}, 13\penalty0 (4):\penalty0 046005, 2018.

\bibitem[Asztalos et~al.(2021)Asztalos, Dawson, and
  Williams]{asztalos2021modeling}
K.~J. Asztalos, S.~T.~M. Dawson, and D.~R. Williams.
\newblock Modeling the flow state sensitivity of actuation response on a
  stalled airfoil.
\newblock \emph{AIAA J.}, 59\penalty0 (8):\penalty0 2901--2915, 2021.

\bibitem[Golestanian et~al.(2011)Golestanian, Yeomans, and
  Uchida]{golestanian2011hydrodynamic}
Ramin Golestanian, Julia~M Yeomans, and Nariya Uchida.
\newblock Hydrodynamic synchronization at low reynolds number.
\newblock \emph{Soft Matter}, 7\penalty0 (7):\penalty0 3074--3082, 2011.

\bibitem[Kawamura and Tsubaki(2018)]{kawamura2018phase}
Yoji Kawamura and Remi Tsubaki.
\newblock Phase reduction approach to elastohydrodynamic synchronization of
  beating flagella.
\newblock \emph{Phys. Rev. E}, 97\penalty0 (2):\penalty0 022212, 2018.

\bibitem[Herrmann et~al.(2020)Herrmann, Oswald, Semaan, and
  Brunton]{herrmann2020modeling}
B.~Herrmann, P.~Oswald, R.~Semaan, and S.~L. Brunton.
\newblock Modeling synchronization in forced turbulent oscillator flows.
\newblock \emph{Comm. Phys.}, 3\penalty0 (1):\penalty0 195, 2020.

\bibitem[Giannenas et~al.(2022)Giannenas, Laizet, and
  Rigas]{giannenas2022harmonic}
A.~E. Giannenas, S.~Laizet, and G.~Rigas.
\newblock Harmonic forcing of a laminar bluff body wake with rear pitching
  flaps.
\newblock \emph{J. Fluid Mech.}, 945:\penalty0 A5, 2022.

\bibitem[Guevara and Glass(1982)]{guevara1982phase}
M.~R. Guevara and L.~Glass.
\newblock Phase locking, period doubling bifurcations and chaos in a
  mathematical model of a periodically driven oscillator: A theory for the
  entrainment of biological oscillators and the generation of cardiac
  dysrhythmias.
\newblock \emph{J. Math. Bio.}, 14:\penalty0 1--23, 1982.

\bibitem[Granada and Herzel(2009)]{granada2009achieve}
A.~E. Granada and H.~Herzel.
\newblock How to achieve fast entrainment$?$ {The} timescale to
  synchronization.
\newblock \emph{PloS One}, 4\penalty0 (9):\penalty0 e7057, 2009.

\bibitem[Zlotnik et~al.(2013)Zlotnik, Chen, Kiss, Tanaka, and
  Li]{zlotnik2013optimal}
A.~Zlotnik, Y.~Chen, I.~Z. Kiss, H.-A. Tanaka, and Jr-S. Li.
\newblock Optimal waveform for fast entrainment of weakly forced nonlinear
  oscillators.
\newblock \emph{Phys. Rev. Lett.}, 111\penalty0 (2):\penalty0 024102, 2013.

\bibitem[Takata et~al.(2021)Takata, Kato, and Nakao]{takata2021fast}
S.~Takata, Y.~Kato, and H.~Nakao.
\newblock Fast optimal entrainment of limit-cycle oscillators by strong
  periodic inputs via phase-amplitude reduction and floquet theory.
\newblock \emph{Chaos: An Inter. J. Nonlin. Sci.}, 31\penalty0 (9):\penalty0
  093124, 2021.

\bibitem[Strogatz(1994)]{strogatz1994chaos}
S.~Strogatz.
\newblock \emph{Nonlinear dynamics and chaos}.
\newblock Perseus Books Group, 1994.

\bibitem[Ermentrout and Kopell(1991)]{ermentrout1991multiple}
G~Bard Ermentrout and Nancy Kopell.
\newblock Multiple pulse interactions and averaging in systems of coupled
  neural oscillators.
\newblock \emph{J. Math. Biol.}, 29\penalty0 (3):\penalty0 195--217, 1991.

\bibitem[Roma et~al.(1999)Roma, Peskin, and Berger]{roma1999adaptive}
A.~M. Roma, C.~S. Peskin, and M.~J. Berger.
\newblock An adaptive version of the immersed boundary method.
\newblock \emph{J. Comp. Phys.}, 153\penalty0 (2):\penalty0 509--534, 1999.

\bibitem[Taira and Colonius(2007)]{taira2007immersed}
K.~Taira and T.~Colonius.
\newblock The immersed boundary method: a projection approach.
\newblock \emph{J. Comp. Phys.}, 225\penalty0 (2):\penalty0 2118--2137, 2007.

\bibitem[Kajishima and Taira(2016)]{kajishima2016computational}
T.~Kajishima and K.~Taira.
\newblock \emph{Computational fluid dynamics: incompressible turbulent flows}.
\newblock Springer, 2016.

\bibitem[Chang(1992)]{chang1992potential}
C.-C. Chang.
\newblock Potential flow and forces for incompressible viscous flow.
\newblock \emph{Proc. Royal Soc. London. Series A: Math. Phys. Sci.},
  437\penalty0 (1901):\penalty0 517--525, 1992.

\bibitem[Eldredge and Jones(2019)]{eldredge2019leading}
J.~D. Eldredge and A.~R. Jones.
\newblock Leading-edge vortices: mechanics and modeling.
\newblock \emph{Ann. Rev. Fluid Mech.}, 51:\penalty0 75--104, 2019.

\end{thebibliography}

\end{document}